\documentclass[aoas,MSNbibl,nameyear,rotating,dvips]{arximspdf}
\usepackage{dcolumn}
\usepackage{graphicx}

\doi{10.1214/12-AOAS584} 
\volume{7}
\issue{1}
\pubyear{2013}
\firstpage{319}
\lastpage{341}

\makeatletter
\newcolumntype{d}[1]{D{.}{.}{#1}}
\newcommand{\bbm}[1]{\mathbf{#1}}
\newcommand{\bmm}[1]{\bolds{#1}}
\makeatother

\begin{document}
\begin{frontmatter}

\title{Robust VIF regression with application to variable selection in large data sets}
\runtitle{Robust VIF regression with application to variable selection}

\begin{aug}
\author[A]{\fnms{Debbie J.} \snm{Dupuis}\corref{}\ead[label=e1]{debbie.dupuis@hec.ca}\thanksref{t1}}
\and
\author[B]{\fnms{Maria-Pia} \snm{Victoria-Feser}\ead[label=e2]{maria-pia.victoriafeser@unige.ch}\thanksref{t2}}
\runauthor{D. J. Dupuis and M.-P. Victoria-Feser}
\thankstext{t1}{Supported by the Natural Sciences
and Engineering Research Council of Canada.}
\thankstext{t2}{Supported by Swiss National Science
Foundation Grant 100018-131906.}
\affiliation{HEC Montr\'eal and HEC Gen\`{e}ve}
\address[A]{Department of Management Sciences\\
HEC Montr\'eal\\
3000, chemin de la C\^ote-Sainte-Catherine\\
Montr\'eal (Qu\'ebec) \\
Canada H3T 2A7\\
\printead{e1}} 
\address[B]{Research Center for Statistics\\
HEC Gen\`{e}ve\\
40, bd du Pont d'Arve\\
CH-1211 Gen\`eve\\
Switzerland\\
\printead{e2}}
\end{aug}

\received{\smonth{11} \syear{2011}}
\revised{\smonth{5} \syear{2012}}

%
\begin{abstract}
The sophisticated and automated means of data collection used by an
increasing number of institutions and companies leads to extremely
large data sets. Subset selection in regression is essential when
a huge number of covariates can potentially explain a response
variable of interest.
The recent statistical li\-terature has seen an emergence of new
selection methods that provide some type of compromise between
implementation (computational speed)
and statistical optimality (e.g., prediction error minimization).
Global methods such as Mallows' $C_p$
have been supplanted by sequential methods such as stepwise regression.
More recently, streamwise regression, faster than the former,
has emerged.
A recently proposed streamwise regression approach based on the
variance inflation factor (VIF) is promising, but its least-squares
based implementation makes it susceptible to the outliers inevitable in such
large data sets. This lack of robustness can lead to poor and suboptimal
feature selection. In our case, we seek to predict an individual's educational
attainment using economic and demographic variables. We show how
classical VIF performs this task poorly and a robust procedure
is necessary for policy makers.
This article proposes a robust VIF regression, based on
fast robust
estimators, that inherits all the good properties of classical VIF in the
absence of outliers, but also continues to perform well in their presence
where the classical approach fails.
\end{abstract}

%

\end{frontmatter}

\section{Introduction}\label{sec1}

Data sets with millions of observations and a huge number of variables
are now
quite common, especially in business- and finance-related fields, as
well as
computer sciences, health sciences, etc.
An important challenge is to provide statistical tools and algorithms that
can be used with such data sets. In particular, for regression models,
a first data analysis requires that the number
of potential explanatory variables be reduced to a reasonable and
tractable amount.
Consider $p$ potential explanatory variables
$[1 \enskip x_1 \cdots x_p]^T = \bbm{x}$ and a response variable $y$
observed on $n$ subjects. The classical normal linear model supposes
$y|\bbm{x}\sim N(\bbm{x}^T\bmm{\beta};\sigma^2)$
with slope parameters $\bmm{\beta}=[\beta_0,\beta_1,\ldots,\beta_p]^T$.
The aim is to find a subset of explanatory variables that satisfies a
given criterion
and such that the regression model holds.


The selection criteria are numerous and can be based on
prediction, fit, etc.
The available selection procedures can be broadly classified into three classes
according to their general strategy and, as a result, their
computational speed.
A first class considers all the possible combinations of covariates as
potential models,
evaluates each according to a fixed criterion, and chooses the model
which best suits
the selected criterion. A second class is formed of sequential
selection procedures in which a covariate at a time is entered in (or
removed from)
the model, based on a criterion that can change from one step to the next
and that is computed for all potential variables to enter (or to exit) until
another criterion is reached. Finally, the third class of selection procedures
is also sequential in nature, but each covariate is only considered once
as a potential covariate. For the first class, we find criteria such as the
AIC [\citet{Akai73}], BIC [\citet{Schw78}], Mallows' $C_p$ [\citet{Mall73}],
cross-validation, etc. [see also \citet{Efro04}]. These
methods are not adapted to large data sets since the number of
potential models
becomes too large and the computations are no longer feasible. For the
second class, we find, for example, the classical stepwise regression
which can be considered as a simple algorithm
to compute the estimator of regression coefficients $\bmm{\beta}$ that minimizes
an $l_q$ penalized sum of squared errors
$\|\bbm{y}-\bbm{X}\bmm{\beta} \|_2^2+\lambda_q\|\bmm{\beta}\|_{l_q}$,
with $q=0$ and
$\bbm{X}=[{\bbm{1}\enskip  \bbm{x}_j}]_{j=1,\ldots,p}$
and $\bbm{1}$ a vector of ones, that is,
$\|\bmm{\beta}\|_{l_0}=\sum_{j=1}^p \eta(\beta_j\neq0)$ [see \citet
{LiFoUn11}],
with $\eta(\beta_j\neq0)=1$ if $\beta_j\neq0$ and $0$ otherwise.
Fast algorithms
for stepwise regressions are available, for example, \citet{FoSt04}.
Procedures for the $l_1$ problem are also available,
for example, Lasso/LARS [\citet{EfHaJoTi04}], the
Dantzig Selector [\citet{CaTa07}], or coordinate descent [\citet{FrHaTi10}].
But these algorithms may also become very slow for large data sets, not
only because
all remaining variables are evaluated at each stage, but also because
the penalty
$\lambda_q$ needs to be computed, and often via cross-validation.
The last class is a variation of stepwise regression in which
covariates are
tested sequentially but only \textit{once} for addition to the model.
An example is the \textit{streamwise regression} of \citet{ZhFoStUn06},
which uses the \mbox{$\alpha$-investing} rule [\citet{FoSt08}], is very fast, and
guards against overfitting. An improved streamwise regression approach
was recently
proposed in \citet{LiFoUn11} where a very fast to compute test
statistic based on the
variance inflation factor (VIF) of the candidate variable, given the
currently selected
model, is proposed. The approach takes into account possible
multicollinearity, seeking to
find the best predictive model, even if it is not the most
parsimonious. Comparisons
in \citet{LiFoUn11} establish that the method performs well and is the
fastest available.

Our concern in this paper is to provide model selection tools for the
regression model
that are robust to small model deviations.
As argued in \citet{DuVF11} [see also \citet{RoSt94}], spurious model
deviations
such as outliers can lead to a
completely different, and suboptimal, selected model when a nonrobust criterion,
like Mallows' $C_p$ or the VIF, is used.
This happens because under slight data contamination, the estimated model
parameters, using, for example, the least squares estimator (LS) and,
consequently,
the model choice criterion,
can be seriously biased. The consequence is that when the estimated criteria
are compared to an absolute level (like a quantile of the $\chi^2$
distribution),
the decisions are taken at the wrong level.
For the first class of selection procedures, robust criteria have been
proposed such as
the robust AIC of \citet{Ronc82}, the robust BIC of \citet{Mach93},
the robust Mallows' $C_p$ of \citet{RoSt94}, and
a robust criterion based on cross-validation (CV) in \citet{RoFiBl97}.
Since standard robust estimators are impossible to compute when the
number of covariates
is too large, \citet{DuVF11} proposed the use of a forward search
procedure together
with adjusted robust estimators when there is a large number of
potential covariates.
Their selection procedure, called
Fast Robust Forward Selection (FRFS), falls in the second class of
selection procedures. FRFS outperforms classical approaches such as Lasso/LARS
when data contamination is present and
outperforms, in all studied instances,
a robust version of the LARS algorithm proposed by
\citet{KhVaZa07}.\looseness=1

However, although FRFS is indeed very fast and robust, it too can
become quite slow
when the number of potential covariates is very large, as \textit{all}
covariates are reconsidered
after one is selected for entry in the model.
It is therefore important to have a robust selection procedure
in the streamwise regression class so that very large data sets can be analyzed
in a robust fashion. In this paper
we develop a robust VIF approach that is fast, very efficient,
and clearly outperforms nonrobust VIF in the presence of outliers.

The remainder of the paper is organized as follows. In Section~\ref{SecFast-algo} we
review the
classical VIF approach and present our robust VIF approach. A
simulation study
in Section~\ref{Secsimulation} shows the good performance of the new approach. In Section~\ref{SecExamples}
we analyze educational attainment data and show how policy makers are
better served by
robust VIF regression than by classical VIF or Lasso.
In Section~\ref{sec5} we present a shorter analysis of a large crime data set
that highlights more problems with classical VIF for real data.
Section~\ref{sec6} contains a few closing remarks.

\section{Robust VIF regression}
\label{SecFast-algo}

\subsection{The classical approach}

Lin, Foster and Ungar (\citeyear{LiFoUn11}) propose a procedure that allows one to sweep through
all available covariates
and to enter those that can reduce a statistically sufficient part of
the variance in
the predictive model.
Let $\bbm{X}_S$ be the design matrix that includes the selected
variables at a given stage, and
$\tilde{\bbm{X}}_S=[\bbm{X}_S\enskip \bbm{z}_j]$ with $\bbm
{z}_j$ the new potential
covariate to be considered for inclusion.
Without loss of generality, we suppose all variables have been standardized.
Consider the following two models:
%
%
\begin{eqnarray}
\bbm{y}&=&\bbm{X}_S\bmm{\beta}_S+\bbm{z}_j
\beta_j + \bmm{\varepsilon}_{\mathrm{step}},\qquad \bmm{
\varepsilon}_{\mathrm{step}}\sim N\bigl(\bbm{0},\sigma^2_{\mathrm{step}}
\bbm{I}\bigr), \label{EqMod-step}
\\
\bbm{r}_S&=&\bbm{z}_j\gamma_j + \bmm{
\varepsilon}_{\mathrm{stage}},\qquad \bmm{\varepsilon}_{\mathrm{stage}}\sim N\bigl(\bbm{0},
\sigma^2_{\mathrm{stage}}\bbm{I}\bigr), \label{EqMod-stage}
\end{eqnarray}
where $\bbm{r}_S=(\bbm{I}-\bbm{X}_S (\bbm{X}_S^T\bbm{X}_S)^{-1}\bbm
{X}_S^T)\bbm{y}$ are the residuals of the projection
of $\bbm{y}$ on $\bbm{X}_S$.
All known estimators of the parameters $\beta_j,\sigma^2_{\mathrm{step}}$ and
$\gamma_j,\sigma^2_{\mathrm{stage}}$ will provide different estimates when the covariates
present some degree of multicollinearity, and, consequently,
significance tests based on
estimates of $\beta_j$ or $\gamma_j$ do not necessarily lead to the
same conclusions.
While in stepwise regression the significance of $\beta_j$ in model
(\ref{EqMod-step})
is at the core of the selection procedure, in streamwise regression
one estimates more conveniently $\gamma_j$.
\citet{LiFoUn11} show that, when LS are used to estimate,
$\hat\gamma_j=\rho\hat\beta_j$ where
$\rho=\bbm{z}_j^T(\bbm{I}-\bbm{X}_S(\bbm{X}_S^T\bbm{X}_S)^{-1}\bbm
{X}_S^T)\bbm{z}_j$. They
then compare $T_\gamma=\hat\gamma_j/(\rho^{1/2}\sigma)$, with suitable
estimates for $\rho$ and $\sigma$, to the standard normal distribution
to decide whether or not $\bbm{z}_j$ should be added to the current model.
The procedure is called VIF regression since \citet{Marq70} called
$1/\rho$ the VIF for $\bbm{z}_j$.

\subsection{A robust weighted slope estimator}

Since the test statistic $T_\gamma$ is based on the following, (1) the
LS estimator $\hat\gamma_j$,
(2) $\rho$, in turn based on the design matrix~$\bbm{X}_S$ and $\bbm
{z}_j$, and (3)
the classical estimator of $\sigma$, it is obviously very sensitive to outliers,
a form of model deviation.
An extreme response or a very badly placed design point can have a
drastic effect
on $T_\gamma$. The latter is then compared to the null distribution:
the correct asymptotic distribution under the hypothesis that the
regression model holds.
With model deviations, the
null distribution is not valid and, hence, selection decisions (to add
the covariate or not)
are taken rather \mbox{arbitrarily}.
We propose here to limit the influence of extreme observations by considering
weighted LS estimators of the form
%
%
\begin{equation}
\label{Eqweighted-LS} \widehat{\bmm{\beta}}=\bigl(\bbm{X}^{wT}
\bbm{X}^w\bigr)^{-1}\bbm{X}^{wT}
\bbm{y}^w,
\end{equation}
with $\bbm{X}^w=\operatorname{diag}(\sqrt{w_{i}^0})\bbm{X}$ and $\bbm{y}^w=\operatorname{diag}(\sqrt {w_{i}^0})\bbm{y}$.
The weights $w_{i}^0$ depend on the data and are such that extreme
observations in
the response and/or in the
design have a nil or limited effect on the value of $\widehat{\bmm{\beta}}$.
\citet{DuVF11} propose Tukey's redescending
biweight weights
%
\begin{equation}
\label{EqTukey-wgt} w_i(r_i;c)= \cases{\displaystyle
\biggl( \biggl(\frac{r_i}{c} \biggr)^2-1
\biggr)^2, & \quad {\mbox{if $|r_i| \leq c$},}
\vspace*{2pt}\cr
0 , &\quad {\mbox{if $|r_i| > c$,}}}
\end{equation}
where $r_i=(y_i-\bbm{x}_i^T\bmm{\beta})/\sigma$ are standardized
residuals that are
computed in practice for chosen estimators of $\bmm{\beta}$ and $\sigma
$ (see below).
The constant $c$ controls the efficiency and the robustness of the
estimator. Indeed, the most
efficient estimator is the LS estimator, that is, (\ref{Eqweighted-LS})
with all weights equal to one (i.e., $c\rightarrow\infty$), but it is
very sensitive to (small) model deviations, while a less efficient but
more robust estimator
is obtained by downweighting observations that have a large influence on
the estimator, that is, by setting $c<\infty$ in (\ref{EqTukey-wgt}).
The value $c=4.685$ corresponds to an efficiency level of 95\% for the
robust estimator compared to the LS estimator at the normal
model and is the value used throughout the paper.

We follow \citet{DuVF11}\vspace*{-1pt} and use for the weights $w_i^0=w_i(r_i^0;c)$
in (\ref{Eqweighted-LS}),
where the residuals $r_i^0=(y_i-\bbm{x}_{i}^T\hat{\bmm{\beta}}{}^0)/\hat
\sigma^0$ and
$\hat\sigma^0=  1.483\operatorname{med}\vert\tilde r_i^0-\operatorname{med}(\tilde r_i^0)
\vert$, the median absolute deviation (MAD) of the residuals
$\tilde{r}_i^0 = y_i-\bbm{x}_{i}^T\widehat{\bmm{\beta}}{}^0$.
The slope estimates are $\widehat{\bmm{\beta}}{}^{0}= [ (\bbm
{X}^w_0)^T\bbm{X}^w_0  ]^{-1}
(\bbm{X}^{w2}_0)^T\bbm{y}$, with $\bbm{X}^w_0 = [
 {1 \enskip \sqrt{w_{i1}}x_{i1} \enskip
\cdots\enskip \sqrt{w_{ip}}x_{ip}}]$
and $\bbm{X}^{w2}_0 = [
{1 \enskip w_{i1}x_{i1} \enskip
\cdots\enskip w_{ip}x_{ip}}]$, $i=1,\ldots,n$, with weights
$w_{ij}$, for all $j=1,\ldots,p$, computed using (\ref{EqTukey-wgt})
at the residuals
$r_{ij}=(y_i-\widehat{\beta}_{0j}-x_{ij}\widehat\beta_j)/\widehat\sigma_j$,
with $\widehat\sigma_j=\operatorname{MAD}(y_i-\widehat{\beta}_{0j}-x_{ij}\widehat
\beta_j)$. The
slope estimators $\widehat\beta_{1},\ldots,\widehat\beta_{p}$ and the
intercept estimators
$\widehat\beta_{01},\ldots,\widehat\beta_{0p}$ are computed on the $p$
marginal models
$y=\beta_{01}+x_{1}\beta_1+\varepsilon_1,\ldots,y=\beta_{0p}+
x_{p}\beta_p+\varepsilon_p$ using a robust weighted estimator defined
implicitly through
%
%
\begin{equation}
\sum_{i=1}^n w_i(r_i;c)r_i
\bbm{x}_{i}=0 . \label{EqM-estim}
\end{equation}
Here we consider Huber's weights given for the regression model by
%
%
\begin{equation}
w_i(r_i;c)= \operatorname{min} \biggl\{1 ; \frac{c}{|r_i|} \biggr\}
\label{EqHuber-wgt},
\end{equation}
with $c=1.345$. Estimators in (\ref{EqM-estim}) belong to the class of
$M$-estimators [\citeauthor{Hube64} (\citeyear{Hube64,Hube67})].
With (\ref{EqHuber-wgt}) in (\ref{EqM-estim}), the marginal intercepts
and slope estimators are simpler (and faster) to
compute than the ones based on Tukey's biweight weights as originally
proposed in \citet{DuVF11}.
For the scale in the weights in (\ref{EqM-estim}), we propose to use
the MAD of the residuals.

The estimator in (\ref{Eqweighted-LS}) is a one-step estimator that is actually
biased when there is multicollinearity in the covariates.
\citet{DuVF11} show that the bias can be made
smaller and even nil if $\widehat{\bmm{\beta}}=\widehat{\bmm{\beta
}}{}^{1}$ is
iterated further to get, say,
$\widehat{\bmm{\beta}}{}^{k}$, computed at the updated weights
$w_i^1,\ldots,w_i^{k-1}$ based on the residuals
$r_{i}^{(1)}=(y_i-\bbm{x}_{i}^T\widehat{\bmm{\beta}}{}^{(1)})/\widehat
\sigma^{(1)}, \ldots,
r_{i}^{(k-1)}=(y_i-\bbm{x}_{i}^T\widehat{\bmm{\beta}}{}^{(k-1)})/\widehat
\sigma^{(k-1)}$.
In the simulation study in Section~\ref{Secsimulation}, however, we
find that the bias is very small even with relatively
large multicollinearity, so that in practice there is often no need to
proceed with
this iterative correction.

Finally, $\widehat{\bmm{\beta}}{}^{0}$ is a coordinate-wise robust estimator
and \citet{AlAeYoZa09} show, through the computation
of a generalized version of the influence function [\citeauthor{Hamp68} (\citeyear{Hamp68,Hamp74})]
and different contamination schemes in the multivariate
normal (MVN) setting, that coordinate-wise robust estimators
can be less sensitive to extreme observations when they
occur independently at the univariate level.

\subsection{Robust VIF selection criterion}\label{sec2.3}

Let $\bbm{X}_S^w=\operatorname{diag}(\sqrt{w_{iS}^0})\bbm{X}_S$ be the weighted design matrix
at stage $S$ with, say, $q$ columns (hence $q-1$ covariates), and
$\bbm{z}_j^w=\operatorname{diag}(\sqrt{w_{ij}})\bbm{z}_j$ the new
candidate covariate that is evaluated at the current stage $S+1$.
One could use the weights $w_{iS}^0$ for $\bbm{z}_j^w$ instead of the
weights $w_{ij}$ computed at the marginal models with only $\bbm{z}_j$
as a covariate, but this would require more computational time. The
simulation results
in Section~\ref{Secsimulation} show that one gets very satisfactory
results with $w_{ij}$.
Let also $\tilde{\bbm{X}}_S^w=[\bbm{X}_S^w | \bbm{z}_j^w]$ and
define $ \widehat{\beta}_{j}^w$ as the last element of the vector
$[\tilde{\bbm{X}}_S^{wT} \tilde{\bbm{X}}_S^w]^{-1} \tilde{\bbm
{X}}_S^{wT} \bbm{y}^w $ with $\bbm{y}^w=\operatorname{diag}(\sqrt{w_{iS}^0})\bbm{y}$.
$ \widehat{\beta}_{j}^w$ is actually a robust estimator of $\beta_j$ in
(\ref{EqMod-step}).
Let $\bbm{H}_S^w=\bbm{X}_S^w(\bbm{X}_S^{wT}\bbm{X}_S^w)^{-1}\bbm
{X}_S^{wT}$ and $\widehat{\bmm{\beta}}_S=(\bbm{X}_S^{wT}\bbm
{X}_S^w)^{-1}\bbm{X}_S^{wT}\bbm{y}^w$, then
\begin{eqnarray*}
\widehat{\beta}_{j}^w&=& -\bigl(\bbm{z}_j^{wT}
\bbm{z}_j^w-\bbm{z}_j^{wT}\bbm
{H}_S^w\bbm{z}_j^w
\bigr)^{-1}\bbm{z}_j^{wT}\bbm{X}_S^w
\bigl(\bbm{X}_S^{wT}\bbm {X}_S^w
\bigr)^{-1}\bbm{X}_S^{wT}\bbm{y}^w
\\
& &{} + \bigl(\bbm{z}_j^{wT}\bbm{z}_j^w-
\bbm{z}_j^{wT}\bbm{H}_S^w\bbm
{z}_j^w\bigr)^{-1}\bbm{z}_j^{wT}
\bbm{y}^w
\\
&=& \bigl(\bbm{z}_j^{wT}\bbm{z}_j^w-
\bbm{z}_j^{wT}\bbm{H}_S^w\bbm
{z}_j^w\bigr)^{-1}\bbm{z}_j^{wT}
\bigl(\bbm{y}^w-\bbm{X}_S^w\widehat{\bmm{
\beta }}_S\bigr)
\\
&=& \bigl(\bbm{z}_j^{wT}\bbm{z}_j^w-
\bbm{z}_j^{wT}\bbm{H}_S^w\bbm
{z}_j^w\bigr)^{-1}\bbm{z}_j^{wT}
\bbm{r}_S^w
\\
&=& \bigl(\bbm{z}_j^{wT}\bbm{z}_j^w-
\bbm{z}_j^{wT}\bbm{H}_S^w\bbm
{z}_j^w\bigr)^{-1}\bigl(\bbm{z}_j^{wT}
\bbm{z}_j^w\bigr) \bigl(\bbm{z}_j^{wT}
\bbm {z}_j^w\bigr)^{-1}\bbm{z}_j^{wT}
\bbm{r}_S^w,
\end{eqnarray*}
where $\bbm{r}_S^w$ are the residuals of the weighted fit of $\bbm
{y}^w$ on $\bbm{X}_S^w$. Let
\[
\rho^w=\bigl(\bbm{z}_j^{wT}
\bbm{z}_j^w\bigr)^{-1}\bigl(
\bbm{z}_{j}^{wT}\bbm {z}_{j}^{w}-
\bbm{z}_{j}^{wT}\bbm{H}_S^w
\bbm{z}_{j}^{w}\bigr),
\]
%
%
then
\[
\widehat{\beta}_{j}^w = \bigl(\rho^{w}
\bigr)^{-1}\widehat{\gamma}_{j}^w,
\]
with $\widehat{\gamma}_{j}^w = (\bbm{z}_j^{wT}\bbm{z}_j^w)^{-1}\bbm
{z}_j^{wT}\bbm{r}_S^w$,
that is, the weighted estimator of the fit of $\bbm{z}_j^w$ on the
weighted residuals $\bbm{r}_S^w$, that is, model (\ref{EqMod-stage}).
Note, however, that $\widehat{\beta}_{j}^w $ is not equal to the last
element of $\widehat{\bmm{\beta}}{}^1_{S+1}$ unless the weights
$w_{iS}^0$ are used for $\bbm{z}_j^w$.
Note also that we can write
\[
\rho^w=1-R^{w2}_{jS},
\]
with
%
%
\begin{equation}
R^{w2}_{jS}=\bbm{z}_{j}^{wT}
\bbm{H}_S^w\bbm{z}_{j}^{w}\bigl(
\bbm {z}_j^{wT}\bbm{z}_j^w
\bigr)^{-1} \label{EqRw2}
\end{equation}
a robust estimate of the coefficient of determination $R^2$. \citet
{ReVF10} propose a robust $R^2$ based on weighted responses and
covariates and (\ref{EqRw2}) is
equivalent to their proposal (with $a=1$, see their Theorem 1) but with
other sets of weights. Moreover, $\rho^w$ is the partial variance of
$\bbm{z}_{j}^{w}$ given $\bbm{X}_S^w$ [see \citet{DuVF11}].

\citet{LiFoUn11} note that using all the data to compute $\rho$ (in
the classical setting) is quite computationally expensive and
they propose a subsampling approach. For the same reason, we also
propose to actually estimate $\rho^w$ by computing (\ref{EqRw2}) on a
randomly chosen subset of size $m=200$.

To derive the $t$-statistic based on $\widehat{\gamma}_{j}^w$, we
follow \citet{LiFoUn11} who base their comparison on the expected
value of the estimated variance of, respectively,
$\widehat\beta_{j}^w$ and $\widehat{\gamma}_{j}^w$. Let $\widehat{\sigma
}_{\mathrm{step}}^2$ and $\widehat{\sigma}_{\mathrm{stage}}^2$ be, respectively, robust
residual variance estimates for models
(\ref{EqMod-step}) and (\ref{EqMod-stage}). Let also $\bbm
{A}_{(i)(j)}$ denote the element $(i,j)$ of matrix $\bbm{A}$. For
$\widehat\beta_{j}^w$, supposing that $w_{ij}/w_i^0\approx1$, we can use
\begin{eqnarray*}
\widehat{\operatorname{Var}\bigl(\hat{\beta}_{j}^w\bigr)}&\approx&
\hat{\sigma }_{\mathrm{step}}^2 \bigl[\tilde{\bbm X}_S^{wT}
\tilde{\bbm X}_S^w \bigr]_{(q+1)(q+1)}^{-1}e_c^{-1}
\\
&=& \hat{\sigma}_{\mathrm{step}}^2\bigl(\bbm{z}_{j}^{wT}
\bbm{z}_{j}^{w}-\bbm {z}_{j}^{wT}
\bbm{H}_S^w\bbm{z}_{j}^{w}
\bigr)^{-1}e_c^{-1}
\\
&=& \hat{\sigma}_{\mathrm{step}}^2\bigl(\rho^w
\bigr)^{-1}\bigl(\bbm{z}_{j}^{wT}\bbm
{z}_{j}^{w}\bigr)^{-1}e_c^{-1}
\\
&=& \frac{ \hat{\sigma}_{\mathrm{step}}^2 } {n} \bigl(\rho^w\bigr)^{-1} \biggl(
\frac
{1}{n}\sum_i\bigl(z_{ij}^{w}
\bigr)^2 \biggr)^{-1}e_c^{-1},
\end{eqnarray*}
with
%
%
\begin{equation}\label{Eqefficiency-ec}
e_c= \biggl[ \int_{-c}^c
\biggl(5 \biggl( \frac{r}{c} \biggr)^{4}-6 \biggl(
\frac{r}{c} \biggr)^{2}+1 \biggr) \,d\Phi(r)
\biggr]^{2} \Big/ \int_{-c}^c
r^{2} \biggl( \biggl( \frac{r}{c} \biggr)^{2}-1
\biggr)^{4}\,d\Phi(r)\hspace*{-35pt}
\end{equation}
and $\Phi$ the standard normal cumulative distribution [see \citet
{HeCaCoVF09}, equation (3.20)]. For $\hat{\gamma}_{j}^w$,
based on the model with $\bbm{r}_S^w$ as
the response and $\bbm{z}_j^w$ as the explanatory variable (without
intercept), we have
\begin{eqnarray*}
\widehat{\operatorname{Var}\bigl(\hat{\gamma}_{j}^w\bigr)}&
\approx& \hat{\sigma }_{\mathrm{stage}}^2\bigl({\bbm
z}_j^{wT} {\bbm z}_j^w
\bigr)^{-1}\tilde{e}_c^{-1}
\\
&=& \frac{\hat{\sigma}_{\mathrm{stage}}^2}{n} \biggl(\frac{1}{n}\sum_i
\bigl(z_{ij}^{w}\bigr)^2 \biggr)^{-1}
\tilde{e}_c^{-1},
\end{eqnarray*}
with $\tilde{e}_c^{-1}$ the efficiency of a robust slope estimator
computed using Huber's weights relative to the LS, which is not equal
to $e_c^{-1}$, the efficiency of a robust slope estimator computed
using Tukey's weights relative to the LS. We will see below that
the computation of the former\vadjust{\goodbreak} is not needed.
Hence, approximating $\hat{\sigma}_{\mathrm{step}}^2\approx\hat{\sigma
}_{\mathrm{stage}}^2 = \hat{\sigma}^2$, we have
\[
\widehat{\operatorname{Var}\bigl(\hat{\beta}_{j}^w\bigr)}
\approx \bigl( \rho^w\bigr)^{-1}\widehat {\operatorname{Var}\bigl(
\hat{\gamma}_{j}^w\bigr)} (e_c/
\tilde{e}_c)^{-1} .
\]
An honest approximate robust test statistic $T_w$ is then given by
\[
\frac{\hat{\beta}_{j}^w}{\sqrt{\operatorname{Var}(\hat{\beta}_{j}^w)}} \approx  \frac{(\rho^{w})^{-1}\hat{\gamma}_{j}^w}{\sqrt{( \rho^w)^{-1}\widehat
{\operatorname{Var}(\hat{\gamma}_{j}^w)} (e_c/\tilde{e}_c)^{-1} }},
\]
that is,
%
\begin{equation}
T_w= \bigl(\rho^{w}\bigr)^{-1/2}
\frac{\hat{\gamma}_{j}^w}{\sqrt{ {\hat
{\sigma}^2}/{n} ({1}/{n}\sum_i z_{ij}^{w2}  )^{-1} e_c^{-1} }}, \label{Eqt-VIF}
\end{equation}
with $\hat{\sigma}^2$ a robust mean squared error for the model with
$\bbm{r}_S^w$
as response and $\bbm{z}_j^w$ as explanatory variable [i.e., model (\ref
{EqMod-stage})]. We use
$\widehat{\sigma} = \operatorname{MAD}(\bbm{r}_S^w - \bbm{z}_{j}^{w}(\bbm
{z}_j^{wT}\bbm{z}_
j^w)^{-1} \bbm{z}_{j}^{wT} \bbm{r}_S^w)$.

Our \textit{fast robust evaluation procedure} is summarized by the
following five steps.
Suppose that we are at stage $S$ and a set of $q-1$ covariates has been
chosen in the model. We are considering
covariate $\bbm{z}_j$ for possible entry. We are working with
$c=4.685$ and
have computed $e_c$ and the weights $w_{ij}$ and~$w_{iS}^0$:
\begin{longlist}[(1)]
\item[(1)] Obtain the residuals $\bbm{r}_S^w = \bbm{y}^w -
\bbm{X}_S^w (\bbm{X}_S^{wT}\bbm{X}_S^w)^{-1}\bbm{X}_S^{wT}\bbm{y}^w$.
\item[(2)] Set $\bbm{z}_j^w=\operatorname{diag}(\sqrt{w_{ij}})\bbm{z}_j$.
Compute $\widehat{\gamma}_{j}^w = (\bbm{z}_j^{wT}\bbm{z}_j^w)^{-1}\bbm{z}_j^{wT}
\bbm{r}_S^w$ and $\widehat{\sigma} =
\operatorname{MAD}(\bbm{r}_S^w - \bbm{z}_{j}^{w}(\bbm{z}_j^{wT}\bbm{z}_j^w)^{-1}
\bbm{z}_{j}^{wT} \bbm{r}_S^w)$.\vspace*{1pt}
\item[(3)] Sample a small subset $\mathcal{I} = \lbrace i_1, \ldots, i_m
\rbrace
\in\lbrace1, \ldots, n \rbrace$ of the observations and let
$\null_{\mathcal{I}}\bbm{x}$ denote the corresponding subsample from the
regressor $\bbm{x}$.
\item[(4)] Let $\null_{\mathcal{I}}\bbm{H}_S^w
=\null_{\mathcal{I}}\bbm{X}_S^w(\null_{\mathcal{I}}\bbm{X}_S^{wT}
\null_{\mathcal{I}}\bbm{X}_S^w)^{-1}\null_{\mathcal{I}}\bbm{X}_S^{wT}$,
compute
$R^{w2}_{jS} = \null_{\mathcal{I}}\bbm{z}_{j}^{wT}
\null_{\mathcal{I}}\bbm{H}_S^w\null_{\mathcal{I}}\bbm{z}_{j}^{w}\times
(\null_{\mathcal{I}}\bbm{z}_j^{wT}\null_{\mathcal{I}}\bbm{z}_j^w)^{-1}$,
and find $\rho^w=1-R^{w2}_{jS}$.
\item[(5)] Compute the approximate $t$-ratio $T_w =
(\rho^{w})^{-1/2}{\hat{\gamma}_{j}^w}/{\sqrt{ {\hat{\sigma}^2}
(\sum_i z_{ij}^{w2}  )^{-1} e_c^{-1} }}$ and compare it to
an adapted quantile to decide whether or not to add $\bbm{z}_j$ to the
current set.
\end{longlist}
A more detailed algorithm in which the decision rule (whether or not to
add the new variable)
is also specified is given in the \hyperref[app]{Appendix}.
Note that in Step~5 above, the rejection quantile, or corresponding probability
$\alpha_j$, is adapted at each step $j$ so that
$\alpha_j$ increases/decreases if a rejection is made/not made.
As explained in \citet{LiFoUn11}, one can think of $\alpha_j$ as a
gambler's wealth and the game is over when $\alpha_j \leq0$.

\subsection{Comparison with the robust $t$-statistic of FRFS}

The $t$-statistic proposed by \citet{DuVF11} [equation (5)] and used
to test whether a candidate covariate is entered in the current\vadjust{\goodbreak} model
can be written as
\[
T^2=\frac{1}{{\sigma}^2\rho^w}\frac{n}{\sum w_{ij}}e_c \bbm
{y}_j^{wT}\bbm{z}_j^w\bigl(
\bbm{z}_j^{wT}\bbm{z}_j^w
\bigr)^{-1}\bbm{z}_j^{wT}\bigl(\bbm {I}-
\bbm{H}_S^w\bigr)\bbm{y}_j^w
\]
with $\bbm{y}_j^w=\operatorname{diag}(\sqrt{w_{ij}})\bbm{y}$. Supposing that $\bbm
{y}_j^w\approx\bbm{y}^w$ and $n/\sum w_{ij}\approx1$, then
%
%
\begin{eqnarray}
\label{Eqt-FRFS}
T^2 &\approx& \frac{1}{{\sigma}^2\rho^w}e_c
\bbm{y}^{wT}\bbm{z}_j^w\hat {
\gamma}_j^w\nonumber
\\
&=& \frac{(\hat{\gamma}_j^w)^2}{{\sigma}^2\rho^{w}(\bbm{z}_{j}^{wT}\bbm
{z}_{j}^{w})^{-1}} e_c \frac{1}{\hat{\gamma}_j^w}
\bbm{y}^{wT}\bbm {z}_j^w\bigl(
\bbm{z}_{j}^{wT}\bbm{z}_{j}^{w}
\bigr)^{-1}
\\
&=& \frac{(\hat{\gamma}_j^w)^2}{{\sigma}^2\rho^{w}(\bbm{z}_{j}^{wT}\bbm
{z}_{j}^{w})^{-1}} e_c \frac{\bbm{y}^{wT}\bbm{z}_j^w}{\bbm
{z}_j^{wT}(\bbm{I}-\bbm{H}_S^w)\bbm{y}^w} .\nonumber
\end{eqnarray}
Hence, $T_w^2$ in (\ref{Eqt-VIF}) and $T^2$ in (\ref{Eqt-FRFS})
differ by
a multiplicative factor of
\begin{eqnarray*}
\kappa= \frac{\bbm{y}_j^{wT}\bbm{z}_j^w}{\bbm{z}_j^{wT}(\bbm{I}-\bbm
{H}_S^w)\bbm{y}_j^w},
\end{eqnarray*}
which is the ratio of the robustly estimated covariance between
$\bbm{z}_j$ and $\bbm{y}$, and the robustly estimated partial
covariance between
$\bbm{z}_j$ and $\bbm{y}$ given $\bbm{X}_S$.
One can notice that in the orthogonal case (and standardized
covariates), we have
$\bbm{z}_j^{wT}\bbm{H}_S^w\approx0$ so that $\kappa\approx1$.
The value of $\kappa$ was
computed in some of the simulations outlined in the following section.
While $\kappa$ maintained a median value of 1 when aggregating over
the 200 simulated data sets at a given setting, its variability
changed with the theoretical $R^2$ and the absence or presence of outliers.
For example, the interquartile range went from a value
near 0 for $R^2=0.20$ and no outliers,
to 5 for $R^2=0.80$ and 5\% outlying responses with high leverage in
the $p=100$ case. There can thus be a considerable difference in the
two test statistics.

\section{Simulation study}
\label{Secsimulation}

We carry out a simulation study to assess the effectiveness
of the model selection approaches outlined above.
First, we create a linear model
%
\begin{equation}
y = X_1 + X_2 + \cdots+ X_k + \sigma
\varepsilon, \label{lm}
\end{equation}
where $X_1, X_2, \ldots, X_k$ are
multivariate normal (MVN) with $\mathrm{E}(X_i)=0$,\break \mbox{$\operatorname{Var}(X_i)= 1$},
and $\operatorname{corr}(X_i,X_{j}) = \theta$,
$i \ne j, i,j=1,\ldots,k$, and
$\varepsilon$ an independent standard normal variable. We
choose
$\theta$
to produce a range of theoretical
$R^2 = (\operatorname{Var}(y)-\sigma^2)/\operatorname{Var}(y)$ values for (\ref{lm}) and
$\sigma$ to give $t$ values for our target
regressors of about 6
under normality as in \citet{RoFiBl97}.
The covariates $X_1, \ldots, X_k$ are
our $k$ target covariates.
Let $e_{k+1}, \ldots, e_p$ be independent standard\vadjust{\goodbreak}
normal variables and use the first
$2k$ to give the $2k$ covariates
\begin{eqnarray*}
X_{k+1} & = & X_1 + \lambda e_{k+1},\qquad
X_{k+2} = X_1 + \lambda e_{k+2},
\\
X_{k+3} & = & X_2 + \lambda e_{k+3},\qquad
X_{k+4} = X_2 + \lambda e_{k+4},
\\
& \vdots&
\\
 X_{3k-1} & = & X_k + \lambda e_{3k-1},\qquad
X_{3k} = X_k + \lambda e_{3k};
\end{eqnarray*}
and the final $p-3k$ to give the $p-3k$ covariates
\[
X_i = e_i,\qquad i=3k+1,\ldots,p.
\]
Variables $X_{k+1},\ldots,X_{3k}$ are noise
covariates that are correlated with our
target covariates, and variables $X_{3k+1}, \ldots, X_p$
are independent noise covariates. Note that the covariates
$X_1, \ldots, X_p$ are then relabeled with a random permutation
of~$1:p$ so that the target covariates do not appear in position
$1:k$, but rather in arbitrary positions. This is necessary to
test the effectiveness of the streamwise variable selection, as
covariates considered early on are favored for entry when many covariates
are correlated.

We consider samples
without and with contamination. Samples with no contamination
are generated using $\varepsilon\sim\mathrm{N}(0,1)$. To allow
for 5\% outliers, we generate using
$\varepsilon\sim95\%\mathrm{N}(0,1) +
5\% \mathrm{N}(30,1)$.
These contaminated cases also
have high leverage $X$-values: $X_1,\ldots,X_k \sim\operatorname{MVN}$ as
before, except
$\operatorname{Var}(X_i) = 5$,
$i=1,\ldots, k$.
This represents the most difficult contamination scheme:
large residuals at high leverage points.
We also investigate the less challenging cases of 5\% outlying
in response only and 5\% high leverage only.
We choose
$\lambda=3.18$ so that $\operatorname{corr}(X_1,X_{k+1}) =
\operatorname{corr}(X_1,X_{k+2}) = \operatorname{corr}(X_2,X_{k+3}) =
\cdots= \operatorname{corr}(X_k,X_{3k}) = 0.3$.

In all simulations we simulated $n$ independent samples, with or
without contamination, to use for
variable selection. Then, another $n$ independent samples without
contamination were simulated for
out-of-sample performance testing.
The out-of-sample performance\vspace*{1pt}
was evaluated using the mean sum of squared errors (MSE),
$\sum_{i=n+1}^{2n} (y_i-{\bbm x}_i^T \widehat{\bmm\beta})^2/n$, 
where $\widehat{\bmm\beta}$
is the estimated coefficient determined by the classical and robust
VIF regression selection procedures or FRFS applied to the training set.
Because the true predictors are known, we also compute the
out-of-sample performance measure using the true ${\bmm\beta}$.
Classical VIF selection was carried out using the \texttt{VIF} package
for R and default argument settings. Robust VIF was also implemented
in R and code is available at
\href{http://neumann.hec.ca/pages/debbie.dupuis/publicVIFfncs.R}{http://neumann.hec.ca/pages/debbie.dupuis/}
\href{http://neumann.hec.ca/pages/debbie.dupuis/publicVIFfncs.R}{publicVIFfncs.R}.
FRFS is also implemented in R as outlined in \citet{DuVF11}.

It should be noted that when evaluating the performance of a given
criterion (here a selection procedure), the evaluation measure
should be chosen in accordance with the performance measure [see
\citet{Gnei11}]. In our case, although the data
are generated from contaminated conditional Gaussian models,
the core model is still Gaussian and we wish to find the
model that best predicts the conditional mean response.
Consequently, a suitable performance measure is the expected squared error.
However, when estimating the expected squared error from data, one can
resort to the mean (i.e., the MSE) only if the data are purely issued
from the postulated (core) Gaussian model. If this is not the case, or if
there is no guarantee that this is the case, like, for example,
with real data, then a more robust performance measure such as the
median absolute
prediction error (MAPE) should be chosen. Hence, in the simulations we
use the MSE, while
with real data sets we use the MAPE to estimate the evaluation
measure for the comparison of the variable selection methods.

\begin{figure}[t!]

\includegraphics{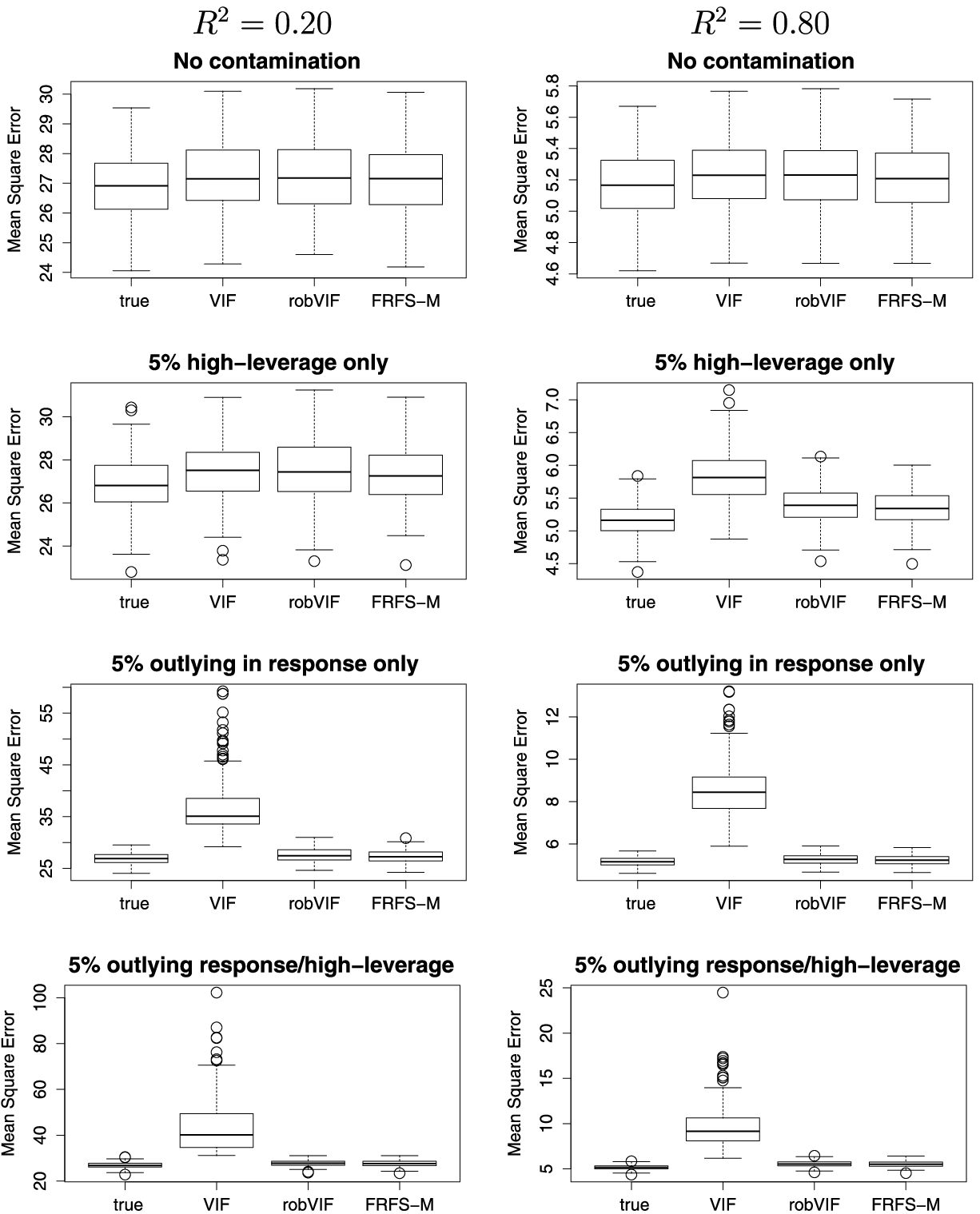}

\caption{Out-of-sample mean square errors of the models chosen by
classical and robust VIF regression, and FRFS-Marginal.
Simulated data, as described in Section \protect\ref{Secsimulation}, have
$n=1000$ observations
with $p=100$ potential regressors, including $k=5$ target
regressors. Correlation among target regressors is
$\theta= 0.1~(R^2=0.20)$ and
$\theta= 0.85~(R^2=0.80)$. Correlation among each target regressor and
two other regressors is 0.3 in all cases. Remaining regressors are
uncorrelated.
Results are based
on 200 simulations for each configuration.} \label{figp100}
\end{figure}

\begin{figure}[t!]

\includegraphics{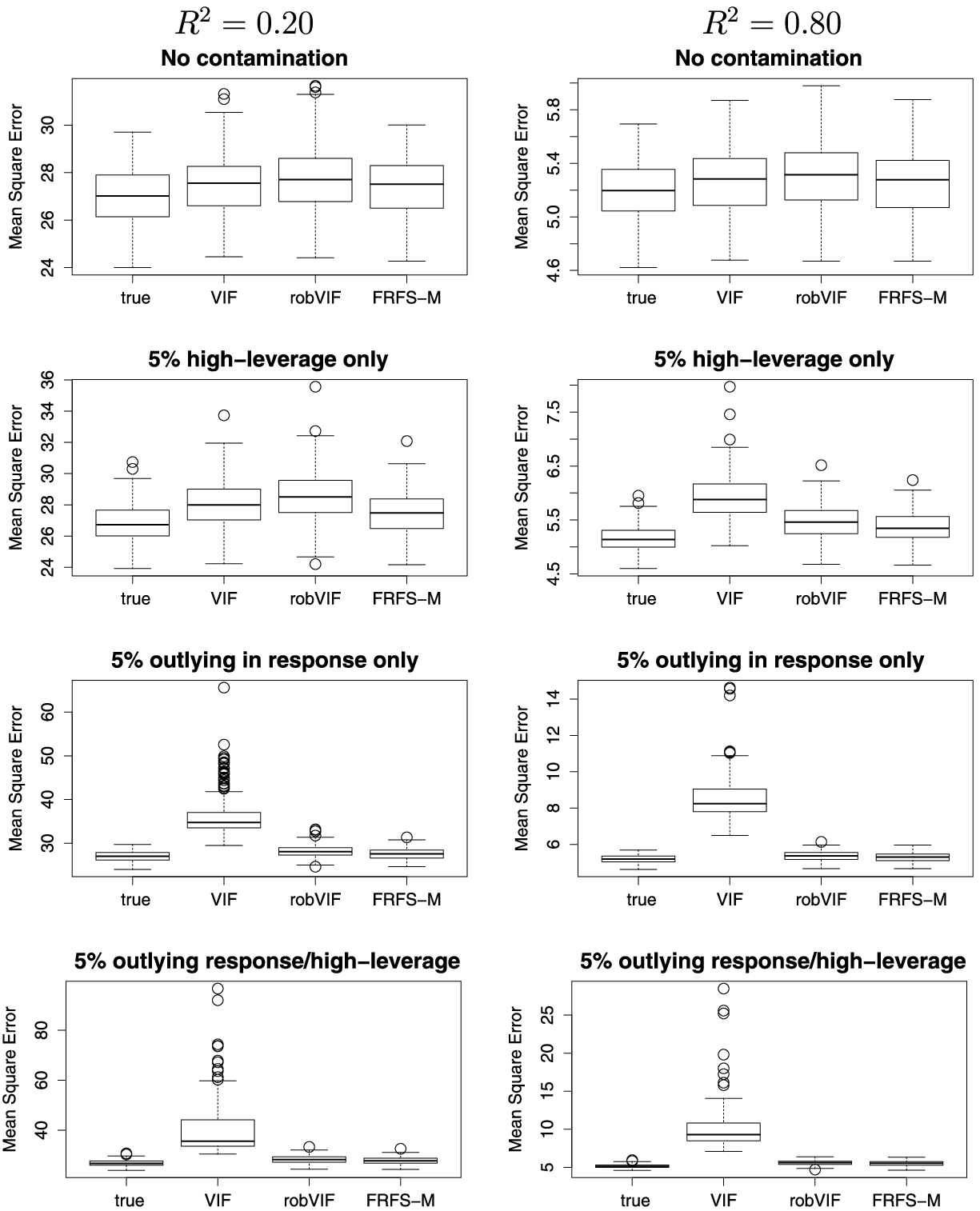}

\caption{Out-of-sample mean square errors of the models chosen by
classical and robust VIF regression, and FRFS-Marginal.
Simulated data, as described in Section \protect\ref{Secsimulation}, have
$n=1000$ observations
with $p=1000$ potential regressors, including $k=5$ target
regressors. Correlation among target regressors is
$\theta= 0.1~(R^2=0.20)$ and
$\theta= 0.85~(R^2=0.80)$. Correlation among each target regressor and
two other regressors is 0.3 in all cases. Remaining regressors are
uncorrelated.
Results are based
on 200 simulations for each configuration.}\label{figp1000}
\end{figure}

Simulations results for $n=1000$, $k=5$, and $p=100$ and $p=1000$,
are presented in Table~\ref{tabk5} and Figures
\ref{figp100} and~\ref{figp1000}, respectively.
Entries in the top panel of the table give the percentage
of runs falling into each category. The category ``Correct'' means
that the correct model was chosen.
``Extra'' means that a model was chosen for which
the true model is a proper subset. ``Missing 1'' means that the
model chosen differed from the true model only in that it was
missing one of the target covariates; ``Missing 2'' and ``Missing~3'' are
defined analogously.
The Monte Carlo standard deviation of entries is bounded by $3.5\%$.
We also report the empirical marginal false discovery rate (mFDR)
$\widehat{\operatorname{mFDR}} = \widehat{E(V)}/(\widehat{E(V)}+\widehat{E(S)}+\eta)$,
where $\widehat{E(S)}$
is the average number of true discoveries, $\widehat{E(V)}$ is the
average number
of false discoveries, and $\eta=10$ is selected following \citet{LiFoUn11}.
We also report the required computation time. Note the particularly frugal
robust approach to VIF regression:
the cost of robustness is no more than a doubling of the
computation time.

%
\begin{sidewaystable}
\tabcolsep=0pt
\tablewidth=\textwidth
\caption{Model selection results.
Simulated data, as described in Section \protect\ref{Secsimulation}, have
$n=1000$ observations
with $p=100$ and $p=1000$ potential regressors, including $k=5$ target
regressors. Correlation among target regressors is
$\theta= 0.1~(R^2=0.20)$ and
$\theta= 0.85~(R^2=0.80)$. Correlation among each target regressor and
two other regressors is 0.3 in all cases. Remaining regressors are
uncorrelated.
Methods are classical (C) and robust (R) VIF regression,
and FRFS-Marginal (F).
Table entries are \% of cases in categories listed in the first column.
Empirical mFDR appears in the second to last row.
Mean execution times (in seconds) appear in the last row.
Data were either not contaminated, had 5\% high leverage
only (hl only), or 5\% outliers
(outlying response and high leverage).
Results are based
on 200 simulations for each configuration}\label{tabk5}
\begin{tabular*}{\textwidth}{@{\extracolsep{\fill}}ld{2.2}d{2.2}d{3.1}d{2.2}d{2.2}d{3.1}d{3.2}d{2.2}d{3.1}d{2.2}d{2.2}d{3.1}d{2.2}d{2.2}d{3.1}d{2.2}d{2.2}d{3.1}@{}}
\hline
&\multicolumn{9}{c}{$\bolds{R^2 = 0.20}$}&\multicolumn{9}{c@{}}{$\bolds{R^2 = 0.80}$}\\[-4pt]
&\multicolumn{9}{c}{\hrulefill}&\multicolumn{9}{c@{}}{\hrulefill}\\
&\multicolumn{3}{c}{\textbf{No contam.}} &
\multicolumn{3}{c}{\textbf{5\% hl only}} &
\multicolumn{3}{c}{\textbf{5\% outliers}} &
\multicolumn{3}{c}{\textbf{No contam.}} &
\multicolumn{3}{c}{\textbf{5\% hl only}} &
\multicolumn{3}{c@{}}{\textbf{5\% outliers}} \\[-4pt]
&\multicolumn{3}{c}{\hrulefill} &
\multicolumn{3}{c}{\hrulefill} &
\multicolumn{3}{c}{\hrulefill} &
\multicolumn{3}{c}{\hrulefill} &
\multicolumn{3}{c}{\hrulefill} &
\multicolumn{3}{c@{}}{\hrulefill} \\
&
\multicolumn{1}{c}{\textbf{C}} &
\multicolumn{1}{c}{\textbf{R}} &
\multicolumn{1}{c}{\textbf{F}} &
\multicolumn{1}{c}{\textbf{C}} &
\multicolumn{1}{c}{\textbf{R}} &
\multicolumn{1}{c}{\textbf{F}} &
\multicolumn{1}{c}{\textbf{C}} &
\multicolumn{1}{c}{\textbf{R}} &
\multicolumn{1}{c}{\textbf{F}} &
\multicolumn{1}{c}{\textbf{C}} &
\multicolumn{1}{c}{\textbf{R}} &
\multicolumn{1}{c}{\textbf{F}} &
\multicolumn{1}{c}{\textbf{C}} &
\multicolumn{1}{c}{\textbf{R}} &
\multicolumn{1}{c}{\textbf{F}} &
\multicolumn{1}{c}{\textbf{C}} &
\multicolumn{1}{c}{\textbf{R}} &
\multicolumn{1}{c@{}}{\textbf{F}} \\
\hline
\multicolumn{19}{c}{$p=100$}\\
\%Correct &13.5 & 33 & 68.5 & 17.5 & 24.5 & 61 &
0 & 20 & 76.5& 11.5&18.5& 86 & 6.5 & 15.5 & 83.5 &
0&15 & 88.5\\
\%Extra &83.5 & 58 & 29.5 & 59.5 & 40 & 24.5 &
0 & 65.0& 20 & 86.5&76.5&12.5& 27.5 & 62.5 & 12.5 &
0&73.5& 7.5\\
\%Missing 1 & 1.5 & 3.5& 1 & 4.5 & 10 & 10 &
0 & 6.5 & 3.5 & 0.5 &1 &1.5 & 7.5 & 5.5 & 4 &
0&3 & 4\\
\%Missing 2 & 0 & 0.5& 0 & 0.5 & 3 & 0 &
1 & 0.5 & 0 & 0 &0 & 0 & 2.5 & 0 & 0 &
1.5&0 & 0\\
\%Missing 3 & 0 & 0 & 0 & 0 & 0 & 0 &
2 & 0 & 0 & 0 &0 & 0 & 0 & 0 & 0 &
11&0 & 0\\
\%Other & 1.5 & 5 & 1 & 18 & 22.5 & 4.5 &
97& 8 & 0 & 1.5 &4 & 0 & 56 & 16.5 & 0 &
87.5&8.5 & 0\\
\%mFDR &11.0 & 6.3& 2.2 & 9.9 & 6.4 & 2.2 &
6.1&9.3 & 1.4 & 16.1&13.2& 0.9& 16 & 13.1 & 1.0 &
10.7&13.8 &0.5\\
Time &0.63 &1.11& 25 & 0.45 & 0.87 & 25 &
0.54&1.09& 25 & 0.69&1.20& 25 & 0.48 & 0.93 & 25 &
0.59&1.20 &25\\ [3pt]
\multicolumn{19}{c}{$p=1000$}\\
\%Correct & 30 & 32 &66.5& 14.5 & 8 & 44 &
0& 25&66.5&14.5& 16& 81 & 1.5 & 8.5 & 69 &
0& 10&72\\
\%Extra & 53 & 27 &26.5& 15.5 & 5 & 15.5 &
0& 26&17.5&77 & 54&12.5& 13.5 & 29 & 12 &
0& 43&9.5\\
\%Missing 1 & 5.5 & 17 & 6 & 25.5 & 24 & 30.5 &
0& 20&12 & 3 & 7.5& 6.5& 5 & 9 & 18 &
0& 6 &18.5\\
\%Missing 2 & 1 & 5.5& 0 & 8.5 & 24 & 3 &
0& 6& 2 & 0 & 0& 0 & 7 & 2 & 1 &
1.5& 1.5& 0\\
\%Missing 3 & 0 & 0.5& 0 & 1.5 & 6.5 & 0 &
0& 1& 0 & 0 & 0& 0 & 0.5 & 0 & 0 &
13& 0 & 0\\
\%Other & 10.5& 18 & 1 & 34.5 & 32.5 & 7 &
100& 22& 2 &5.5 &22.5& 0 & 72.5 & 51.5 & 0 &
85.5& 39.5& 0 \\
\%mFDR & 7.0 & 4.4&2.4 & 5.4 & 3.5 & 2.0 &
4.6& 6.0&1.4 &15.5&13.6&1.0 & 15.6 & 12.9 & 0.9 &
6.2&13.5&0.7\\
Time & 5.8 &10.8&253 & 4.4 & 8.9 & 250 &
6.1&11.7&238 &5.86&10.9&251 & 4.6 & 9.3 & 254 &
5.47&11.3&243 \\
\hline
\end{tabular*}
\end{sidewaystable}

%

%

Both VIF algorithms do not perform well
in terms of the proportion of correctly selected models
and the FRFS-Marginal procedure clearly outperforms
in this respect. The execution time of the
FRFS-Marginal procedure, the faster of
the two FRFS approaches presented in \citet{DuVF11},
is roughly 25 times longer than
that of the robust VIF procedure for these sizes
of data sets.
Both VIF algorithms do, however, choose a model for which the
true model is a subset when there are no outliers.
The classical VIF approach fails miserably in
the presence of outliers (outlying response/high leverage),
while the robust VIF approach
is only slightly affected by the presence of outliers.
The classical VIF approach is less affected by the presence
of high leverage points only, but the effect is increased
under more highly correlated regressors or a higher number
of potential regressors. Results (not shown)
for response variable only
outliers are very similar to outlying response/high leverage
outliers. Finally, other simulations (not shown) reveal that for
less outlying contamination, the robust approaches always
maintain good performance, while the negative impact on
classical VIF is proportional to the level of outlyingness.

As the simulated data sets have noise covariates that
are correlated with target covariates, the poor performance
in terms of \%Correct is expected given
the streamwise approach of VIF regressions.
But as pointed out by \citet{LiFoUn11},
the goal here is different: good fast out-of-sample prediction,
that is, one sacrifices
parsimony for speed.
The streamwise approach is fast and the
main purpose of an $\alpha$-investing control
is to avoid model overfitting. We assess the latter 
through out-of-sample performance.
Figure~\ref{figp100} shows out-of-sample MSE for the
case $p=100$. Robust VIF is as efficient as
classical VIF when there are no outliers (top panel)
and clearly outperforms classical VIF when there is 5\%
contaminated observations
(bottom panels). Robust VIF also loses very little with
respect to FRFS-Marginal. Note that classical VIF seems to offer
some resistance to contamination by high-leverage points only
(as was also seen in Table~\ref{tabk5}), but
completely falls apart in the presence of outlying response
values, and this whether the outlying responses appear at
high-leverage points or not.
Much of the same can be seen in Figure~\ref{figp1000} where
results for the case $p=1000$ are shown.

\section{College data}
\label{SecExamples}
Understanding the factors impacting an individual's educational attainment
is a preoccupation for many governmental and nongovernmental organizations.
For example, a nation's government that recognizes the potential economic
benefits of higher education will seek to write public policies to
promote it.
Private industry that benefits from a well-educated labor market will let
it affect decision making, for example, a company may choose to establish
itself where lifelong education is easily accessible to its personnel.
Finally, an individual's family who associates
personal achievement with higher levels of education may also
act accordingly.

Since the first work by \citet{We76} on projecting community college
enrollments in Arizona, many researchers have sought
to identify the factors impacting educational attainment; see, for
example, \citet{PeMcWi02}, \citet{PeSa02}, \citet{KiAlMe07}, and \citet{Cl11}
(and references therein) for a list of various
studies.

The data analyzed here
are in the R package \texttt{AER} and are a subset of the data
previously analyzed in \citet{Ro95}.
There are 4739 observations on 14 variables. The variables are
listed in Table~\ref{tabcollege1}.
We seek to predict the number
of years of education using 13 economic and demographic variables.
There are continuous and binary variables along with one categorical
variable with three categories which is converted to two dummy variables.
When considering only first-order variables we thus have
$n=4739$ and $p=14$; when we include second-order interaction terms
$p$ rises to 104 (some interaction terms are constant and are removed).
We have standardized the variables.
Our analysis will show how classical, that is, nonrobust,
VIF regression can be inadequate for the policy maker by
failing to keep important features.\looseness=1

%
\begin{table}
\caption{Original 14 variables in college data}\label{tabcollege1}
\begin{tabular*}{\textwidth}{@{\extracolsep{\fill}}ll@{}}
\hline
\multicolumn{1}{@{}l}{\textbf{Variable}} & \textbf{Description} \\
\hline
\texttt{gender} & Factor indicating gender. \\
\texttt{ethnicity} & Factor indicating ethnicity (African-American, Hispanic
or other).\\
\texttt{score} & Base year composite test score. These are achievement
tests given \\
& to high school seniors in the sample.\\
\texttt{fcollege} & Factor. Is the father a college graduate? \\
\texttt{mcollege} & Factor. Is the mother a college graduate? \\
\texttt{home} & Factor. Does the family own their home? \\
\texttt{urban} & Factor. Is the school in an urban area? \\
\texttt{unemp} & County unemployment rate in 1980. \\
\texttt{wage} & State hourly wage in manufacturing in 1980. \\
\texttt{distance} & Distance from 4-year college (in 10 miles). \\
\texttt{tuition} & Average state 4-year college tuition (in 1000 USD). \\
\texttt{income} & Factor. Is the family income above USD 25,000 per year?
\\
\texttt{region} & Factor indicating region (West or other). \\
\texttt{education} & Number of years of education. \\ \hline
\end{tabular*}
\end{table}

The selected models are
compared using the median absolute prediction error (MAPE), as measured
by 10-fold CV. That is, we split the
data into 10 roughly equal-sized parts. For the $k$th part,
we carry out model selection using the other nine parts of the data and
calculate the MAPE of the chosen model when predicting the $k$th
part of the data. We do this for $k=1,\ldots,10$ and show boxplots
of the 10 estimates of the MAPE. For all methods, the data were split
in the same way. For the college data, we randomly generated the folds.
Note here that we look at MAPE instead of mean squared
prediction error, as
these real data can contain outliers (as opposed to the simulated data
which were clean) and the MAPE is a better choice.\looseness=1

%
%
\begin{table}
\caption{VIF and robust VIF
selected variables and estimated slope parameters ($t$-values)\break
when only considering
first-order terms. FRFS-Marginal and\break  FRFS-Full selected
variables and estimated slope parameters ($t$-values) are
also shown. Significance: $\null^{*}$0.05, $\null^{**}$0.01,
$\null^{***}$0.001}\label{tabcollege2}
%
\begin{tabular*}{\textwidth}{@{\extracolsep{\fill}}ld{2.12}d{2.13}d{2.13}@{}}
\hline
\multicolumn{1}{c}{} &
\multicolumn{1}{c}{\textbf{VIF}} &
\multicolumn{1}{c}{\textbf{robVIF}} &
\multicolumn{1}{c}{\textbf{FRFS-Marg/Full}} \\
\textbf{Variable} &
\multicolumn{1}{c}{$\bolds{\widehat\beta}_{\mathbf{LS}}$} &
\multicolumn{1}{c}{$\bolds{\widehat\beta}_{\mathbf{rob}}$} &
\multicolumn{1}{c}{$\bolds{\widehat\beta}_{\mathbf{rob}}$} \\
\hline
\texttt{ethnicityafam} & 0.130\ (5.28)^{***} & 0.129\ (4.90)^{***} &
0.133\ (5.16)^{***} \\
\texttt{ethnicityhispanic}& 0.142\ (5.97)^{***} & 0.124\ (4.92)^{***} &
0.130\ (5.19)^{***} \\
\texttt{score} & 0.772\ (31.3)^{***} & 0.820\ (31.8)^{***} &
0.824\ (31.9)^{***} \\
\texttt{fcollegeyes} & 0.219\ (8.40)^{***} & 0.233\ (8.51)^{***} &
0.232\ (8.52)^{***} \\
\texttt{mcollegeyes} & 0.131\ (5.25)^{***} & 0.146\ (5.60)^{***} &
0.145\ (5.55)^{***} \\
\texttt{homeyes} & 0.054\ (2.39)^{*} & 0.057\ (2.38)^{*} &
0.057\ (2.41)^{*} \\
\texttt{urbanyes} & \multicolumn{1}{c}{--} & 0.024\ (0.96) &
\multicolumn{1}{c}{--} \\
\texttt{unemp} & \multicolumn{1}{c}{--} & 0.077\ (3.00)^{**} &
0.075\ (2.94)^{**} \\
\texttt{wage} & \multicolumn{1}{c}{--} & -0.064\ (-2.56)^{*} &
-0.062\ (-2.50)^{*} \\
\texttt{distance} & -0.064\ (-2.81)^{**} & -0.083\ (-3.22)^{***} &
-0.088\ (-3.57)^{***} \\
\texttt{incomehigh} & 0.163\ (6.70)^{***} & 0.180\ (7.07)^{***} &
0.183\ (7.20)^{***} \\
\texttt{genderfemale} & \multicolumn{1}{c}{--} & \multicolumn{1}{c}{--} &
0.066\ (2.81)^{**} \\
\hline
\end{tabular*}

\end{table}
%

%
\begin{table}
\caption{VIF and robust VIF selected variables and estimated slope
parameters ($t$-values)
when including second-order interactions.
Significance: $\null^{*}$0.05, $\null^{**}$0.01,
$\null^{***}$0.001}\label{tabcollege3}
\begin{tabular*}{\textwidth}{@{\extracolsep{\fill}}ld{2.12}d{2.12}@{}}
\hline
\textbf{Variable} & \multicolumn{1}{c}{$\bolds{\widehat{\beta}}_{\mathbf{LS}}$} & \multicolumn{1}{c@{}}{$\bolds{\widehat{\beta}}_{\mathbf{rob}}$} \\
\hline
\texttt{ethnicityafam} &  0.132\ (5.39)^{***} & 0.127\ (4.83)^{***} \\
\texttt{ethnicityhispanic} & -0.143\ (6.02)^{***} & 0.122\ (4.83)^{***} \\
\texttt{score} &  0.772\ (31.1)^{***} & 0.809\ (27.3)^{***} \\
\texttt{fcollegeyes} &  0.222\ (8.52)^{***} &-0.033\ (-0.17) \\
\texttt{mcollegeyes} &  0.056\ (1.62) & 0.045\ (0.25) \\
\texttt{homeyes} &  0.056\ ( 2.46)^{*} &  0.041\ (1.61) \\
\texttt{urbanyes} & \multicolumn{1}{c}{--} &  0.028\ (1.12) \\
\texttt{unemp} & \multicolumn{1}{c}{--} &  0.059\ (2.10)^{*} \\
\texttt{wage} & \multicolumn{1}{c}{--} & -0.067\ (-2.36)^{*} \\
\texttt{distance} & -0.062\ (-2.75)^{**} & -0.078\ (-3.00)^{**} \\
\texttt{incomehigh} &  0.167\ ( 6.87)^{***} &  0.040\ (0.27) \\
\texttt{genderfemale:score} &  0.030\ (1.24) & \multicolumn{1}{c}{--} \\
\texttt{genderfemale:fcollegeyes} & \multicolumn{1}{c}{--} &  0.002\ (0.06) \\
\texttt{genderfemale:mcollegeyes} & 0.104\ ( 3.07)^{**} & 0.132\ (3.43)^{***} \\
\texttt{score:incomehigh} & \multicolumn{1}{c}{--} & 0.150\ (0.98) \\
\texttt{fcollegeyes:homeyes} & \multicolumn{1}{c}{--} & 0.115\ (1.74) \\
\texttt{fcollegeyes:unemp} & \multicolumn{1}{c}{--} & 0.087\ (1.24) \\
\texttt{fcollegeyes:wage} & \multicolumn{1}{c}{--} & 0.001\ (0.01) \\
\texttt{fcollegeyes:tuition} & \multicolumn{1}{c}{--} & 0.085\ (1.44) \\
\texttt{mcollegeyes:wage} & \multicolumn{1}{c}{--} & 0.002\ (0.01) \\ \hline
\end{tabular*}
\end{table}

For completeness, we compare the models selected by classical and
robust VIF approaches with those of FRFS-Marginal and FRFS-Full where
feasible, as well as that of the popular least angle regression
(LARS) of \citet{EfHaJoTi04}, an extremely efficient algorithm for computing
the entire Lasso [\citet{Tibs96}] path.
We use the R package \texttt{lars}
to do the computations.

Tables~\ref{tabcollege2} and~\ref{tabcollege3} list the VIF and robust
VIF regression selected features, along with estimated slopes,
for the $p=14$ and $p=104$ scenarios, respectively.
For both scenarios, the robust VIF regression approach selects slightly
more, and/or slightly different, features. When considering only first-order
terms, we see that the classical and robust estimates of commonly
selected features are almost the same. This serves as a good form
of validation for the relative importance of these features. However, the
presence of outliers in the data has led classical VIF regression to
completely miss two important features which are identified by robust
VIF regression: \texttt{unemp} and \texttt{wage}. Even LS estimates (not shown)
of the robust VIF regression selected model find these two features important
with $t$-values of 3.15 and $-2.70$, but the classical VIF regression selection
procedure could not detect this importance for the reasons outlined in
the \hyperref[sec1]{Introduction}. FRFS-Marginal and FRFS-Full selected features are identical.
The latter features, along with
estimated slopes, are also shown in Table~\ref{tabcollege2}.

%

%
%
%
%
%

%
\begin{figure}[b]\vspace*{6pt}

\includegraphics{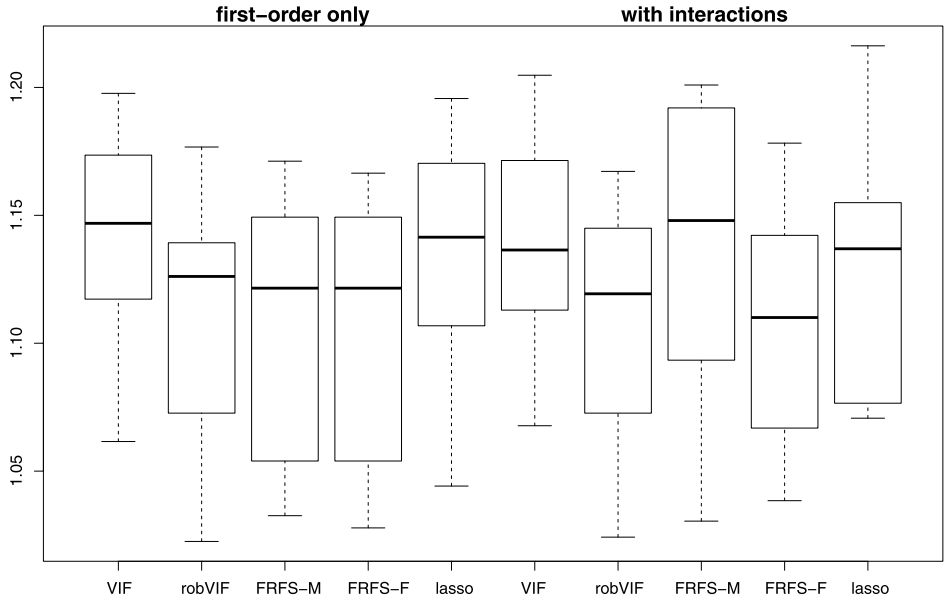}

\caption{College data: Out-of-sample median absolute prediction errors
of the models chosen by
classical and robust VIF regression,
FRFS-Marginal, FRFS-Full and the Lasso,
in 10-fold cross-validation.} \label{Figcollege}
\end{figure}

VIF regression also misses the two important
features in the $p=104$ scenario; see Table~\ref{tabcollege3}.
As both the county unemployment rate and the state
hourly wage in manufacturing are directly impacted by economic policy,
policy makers must be equipped with the best feature selection tools to
have an effective strategy to reach sought after goals: in this case,
increasing the level of education among its constituents.
These tools, we argue, must include a robust selection procedure, as shown
effectively by this example. Further evidence is given in
Figure~\ref{Figcollege} where MAPE for VIF, robust VIF, FRFS-Marginal,
and FRFS-Full and Lasso are shown for
both scenarios. Robust VIF outperforms both of its nonrobust competitors,
and even does better than FRFS-Marginal in the highly collinear case
including interactions. It was shown in \citet{DuVF11} that FRFS-Marginal
could select too few features in the highly collinear case and this motivated
the development of FRFS-Full therein.

%
\begin{table}
\caption{Number of variables selected by VIF and robust VIF
in 100 analyses of college data, each analysis with covariates
presented in a random order}\label{tabcollegerandom}
\begin{tabular*}{\textwidth}{@{\extracolsep{\fill}}lcccccccc@{}}
\hline
\multicolumn{1}{@{}l}{\textbf{\# selected}} & \textbf{7} & \textbf{8} & \textbf{9} &\textbf{10} &\textbf{11} &\textbf{12} &\textbf{13} &\textbf{14}\\
\hline
VIF & 11 &29 &24 &22 &10 &\phantom{0}3 &\phantom{0}1 &-- \\
robVIF & \phantom{0}4 &13 & \phantom{0}8 &23 &17 &12 &10 &13 \\
\hline
\end{tabular*}
\end{table}

%
\begin{table}[b]
\caption{Number of analyses where variable was selected by VIF and
robust VIF
in 100 analyses of college data, each analysis with covariates
presented in a random order}\label{tabcollegerandom2}
\begin{tabular*}{\textwidth}{@{\extracolsep{\fill}}lcc@{}}
\hline
\multicolumn{1}{@{}l}{\textbf{Variable}} & \textbf{VIF} & \textbf{robVIF} \\
\hline
\texttt{genderfemale} & \phantom{0}43 & \phantom{0}47 \\
\texttt{ethnicityafam} & 100 & 100 \\
\texttt{ethnicityhispanic} & \phantom{0}67 & \phantom{0}73 \\
\texttt{score} & 100 & 100 \\
\texttt{fcollegeyes} & 100 & \phantom{0}99 \\
\texttt{mcollegeyes} & 100 & 100 \\
\texttt{homeyes} & \phantom{0}79 & \phantom{0}94 \\
\texttt{urbanyes} & \phantom{00}3 & \phantom{0}38 \\
\texttt{unemp} & \phantom{0}24 & \phantom{0}54 \\
\texttt{wage} & \phantom{0}31 & \phantom{0}63 \\
\texttt{distance} & 100 & \phantom{0}98 \\
\texttt{tuition} & \phantom{0}26 & \phantom{0}56 \\
\texttt{incomehigh} & 100 & \phantom{0}98 \\
\texttt{regionwest} & \phantom{0}31 & \phantom{0}57 \\
\hline
\end{tabular*}
\end{table}

\begin{figure}

\includegraphics{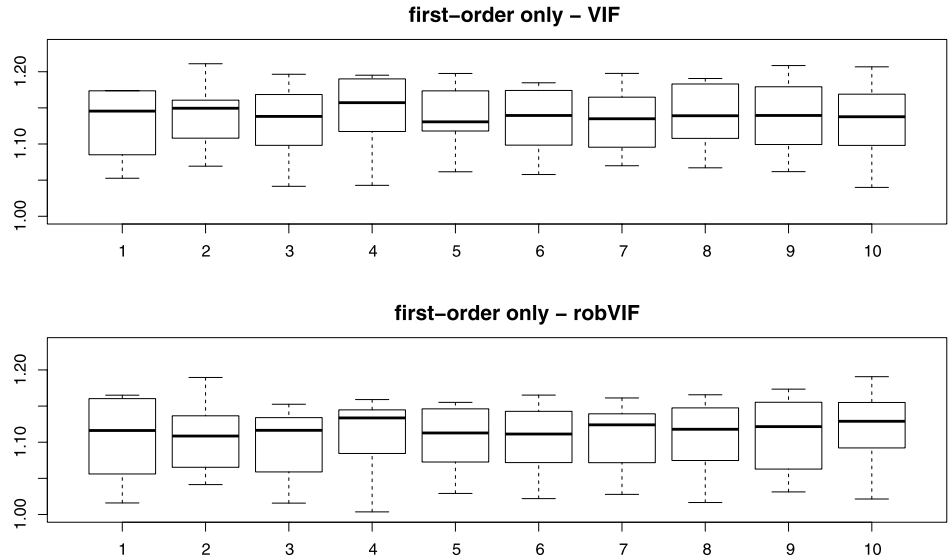}

\caption{College data: Out-of-sample MAPE
of 10 random chosen analyses among 100 analyses reported
in Tables \protect\ref{tabcollegerandom} and \protect\ref{tabcollegerandom2}
for classical and robust VIF regression.
MAPE calculated based on 10-fold cross-validation.}\label{Figcollegerandom}
\end{figure}

As the solution for VIF and robust VIF regression can depend on the
order of the covariates, we ran each procedure several times with
the covariates presented in random order to investigate the stability
of the selected models in terms of model size and prediction performance.
Table~\ref{tabcollegerandom} shows the distribution of the size of
the selected model over 100 analyses and Table~\ref{tabcollegerandom2}
shows how often each variable was selected over these 100 analyses.
As expected, there is considerable variability in the size of the
model, and this both in the classical and robust approaches.
We see, however, that the dominating features are nearly always present.
Note also that \texttt{unemp} and \texttt{wage} are selected twice as often
in the robust approach compared to the classical approach.
In terms of prediction performance, we see in Figure
\ref{Figcollegerandom}\ that the variability in the latter is considerably
less, each of the 10 random analyses shown yielding more or
less the same prediction performance despite the differences
in terms of selected model size and features.

%

\section{Crime data}\label{sec5}
In this section we present a shorter analysis of another data set to
show how the classical approach can even fail to give a usable result.
Also, by looking at a considerably larger data set we can show how
robust VIF provides robust prediction where no other robust method
is feasible.

\begin{figure}

\includegraphics{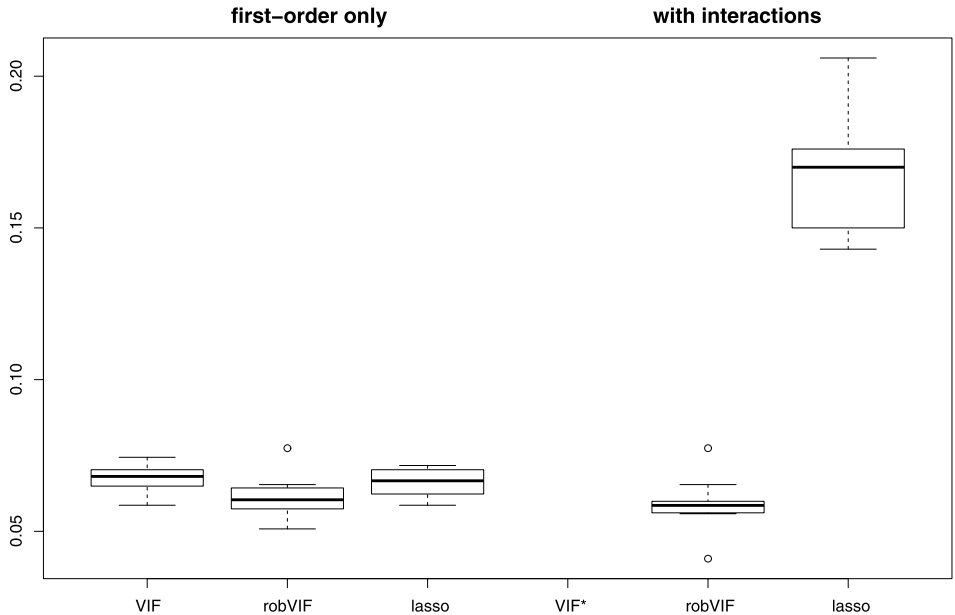}

\caption{Crime and communities data:
Out-of-sample median absolute prediction errors
of the models chosen by
classical and robust VIF regression, and the Lasso,
in 10-fold cross-validation. $^*$Results are not shown
as VIF collapses in 4 folds, yielding MAPE
of 5.62, 6.55, 6.82, 9.41, and 15.1, respectively.
Results for other folds were good, 0.0652, 0.0676, 0.0686, 0.0694,
0.0744, but are excluded from the boxplot to
allow for a better comparisons of all
methods.}\label{FigMAPE-Crime}\vspace*{6pt}
\end{figure}

We analyze recently made available crime data.
These data are from the UCI Machine Learning Repository [\citet{FrAs10}]
and are
available at \href{http://archive.ics.uci.edu/ml/datasets/Communities+and+Crime}{http://archive.ics.uci.edu/ml/datasets/Communities+and+Crime}.
We seek to predict the per capita violent crimes rate using economic,
demographic, community, and law enforcement related variables. After
removing variables with missing data, we are left with $n=1994$ observations
on $p=97$ first-order covariates. If we include second-order
interactions (removing those
that are constant), we have $p=4753$. In both cases, we standardized
the variables.
VIF regression selects 33 and 1437 variables, in the respective scenarios,
while robust VIF regression selects 20 variables in both cases.
Classical
VIF experiences problems with the larger data set, which contains outliers
in a highly multicollinear setting, and chooses too many covariates.
This shows
how the guarantee of no overfitting only holds at the model, that is, without
any outliers in the data.
For these data, robust VIF regression provides the only viable option for
policy makers, as the 1437 features returned by classical VIF regression
do not provide useful information.
As can be seen in Figure~\ref{FigMAPE-Crime}, robust VIF is clearly
the best
performer for both scenarios. VIF regression chooses too many features
for many of the folds and this leads to catastrophic results out-of-sample.

%

\section{Concluding remarks}\label{sec6}

In \citet{LiFoUn11} it was also shown that classical VIF regression
equates or outperforms \textit{stepwise regression}, \textit{Lasso},
\textit{FoBa}, an adaptive forward-backward greedy algorithm focusing
on linear models [\citet{Zhan09}], and \textit{GPS}, the generalized
path-seeking algorithm of \citet{Frie08}. In this paper we present
a very efficient robust VIF approach that clearly outperforms
classical VIF in the case of contaminated data sets.
This robust implementation comes
with a very small cost in speed, computation time is less than doubled,
and provides a much-needed robust model selection for large data
sets.

\begin{appendix}\label{app}

\section*{Appendix: Algorithm robust VIF regression}

The robust VIF regression procedure, based on a streamwise regression
approach and
$\alpha$-investing, can be summarized by the following algorithm:

\begin{description}
\item[Input] data ${\bbm y}, {\bbm x}_1, {\bbm x}_2, \ldots$ (standardized)
\item[Set] initial wealth $a_0=0.50$, pay-out $\Delta a=0.05$,
subsample size $m$,
and robustness constant $c$
\item[Compute] efficiency $e_c^{-1}$ where $e_c$ is as in (\ref
{Eqefficiency-ec})
\item[Get] all marginal weights $w_{ij}$ by fitting $p$ marginal models
$y=\beta_{01}+x_{1}\beta_1+\varepsilon_1,\ldots,y=\beta_{0p}+
x_{p}\beta_p+\varepsilon_p$ using (\ref{EqM-estim}) and (\ref{EqHuber-wgt})
\item[Initialize] $j=1$, $S=\lbrace0 \rbrace$, ${\bbm X}_S = {\bbm1}$,
$\bbm{X}_S^w=\operatorname{diag}(\sqrt{w_{iS}^0})\bbm{X}_S$ and
$\bbm{y}^w=\operatorname{diag}(\sqrt{w_{iS}^0})\bbm{y}$
where
$w_{iS}^0$ is computed using (\ref{EqTukey-wgt}) where
$\bbm{r}^0=(\bbm{y}-\bbm{1}\hat{\bmm{\beta}}{}^0)/\hat\sigma^0$
using
$\bbm{X}^w_0 = \bbm{X}^{w2}_0 = {\bbm1}$,
$\widehat{\bmm{\beta}}{}^{0}= [ (\bbm{X}^w_0)^T\bbm{X}^w_0  ]^{-1}
(\bbm{X}^{w2}_0)^T\bbm{y}$,
where
$\hat\sigma{}^0= 1.483\operatorname{med}\vert\tilde{\bbm{r}}{}^0-\operatorname{med}(\tilde
{\bbm r}{}^0)\vert$
and
$\tilde{\bbm{r}}{}^0 = \bbm{y}-\bbm{1}\widehat{\bmm{\beta}}{}^0$.
\item[repeat]
\begin{description}
\item[ ]
\item[set] $\alpha_j = a_j/(1+j-f)$
\item[get] $T_w$ from the five-step \textit{Fast Robust Evaluation
Procedure} in Section~\ref{sec2.3}.
\item[if] $2(1-\Phi(|T_w|)) < \alpha_j$ \textbf{then}
\[
S = S \cup\lbrace j \rbrace,\hspace*{-3pt}\quad {\bbm X}_S = [ {\bbm1} \enskip{\bbm
x}_j ],\hspace*{-1pt}\quad \bbm{X}_S^w=\operatorname{diag}\Bigl(\sqrt
{w_{iS}^0}\Bigr)\bbm{X}_S,\quad\hspace*{-3pt}
\bbm{y}^w=\operatorname{diag}\Bigl(\sqrt{w_{iS}^0}
\Bigr)\bbm{y},
\]
where
$w_{iS}^0$ is computed using\vspace*{1pt} (\ref{EqTukey-wgt}) where
$\bbm{r}^0=(\bbm{y}-{\bbm X}_S \hat{\bmm{\beta}}{}^0)/\hat\sigma^0$
using
$\bbm{X}^w_0 = [ {1\enskip \sqrt{w_{ij}} x_{ij}} ]$,\vspace*{1pt}
$\bbm{X}^{w2}_0 = [ {1 \enskip w_{ij} x_{ij}} ]$, $i=1,\ldots,n$,
$\widehat{\bmm{\beta}}{}^{0}= [ (\bbm{X}^w_0)^T\bbm{X}^w_0  ]^{-1}
(\bbm{X}^{w2}_0)^T\bbm{y}$,
where
$\hat\sigma^0= 1.483\operatorname{med}\vert\tilde{\bbm{r}}{}^0-\operatorname{med}(\tilde
{\bbm r}{}^0)\vert$
and
$\tilde{\bbm{r}}{}^0 = \bbm{y}-{\bbm X}_S\widehat{\bmm{\beta}}{}^0$.

$a_{j+1} = a_j + \Delta a$

$f = j$
\item[else] $a_{j+1} = a_j - \alpha_j/(1-\alpha_j)$
\item[end if]
\item[{$j=j+1$}]
\end{description}
\item[until] all $p$ covariates have been considered.
\end{description}

\end{appendix}

\section*{Acknowledgments}
The authors thank the Editor and three
referees for comments that improved the presentation.

%


\printaddresses

\end{document}